\documentclass[aps,pra,amsfonts,amssymb,amsmath,showpacs,
floatfix,nofootinbib,groupedaddress,superscriptaddress,citesort,twocolumn]{revtex4}

\usepackage{amssymb,amsmath,amsthm}

\theoremstyle{definition}

\theoremstyle{remark}

\usepackage{subfigure}
\usepackage{mathrsfs}
\usepackage{longtable}
\usepackage{amsfonts}
\usepackage{amstext}
\usepackage[usenames]{xcolor}
\usepackage[dvips]{graphicx}
\def\qed{\leavevmode\unskip\penalty9999 \hbox{}\nobreak\hfill
     \quad\hbox{\leavevmode  \hbox to.77778em{%
              \hfil\vrule   \vbox to.675em%
               {\hrule width.6em\vfil\hrule}\vrule\hfil}}
     \par\vskip3pt}

\begin{document}
\title{Comment on ``Optimal convex approximations of quantum states" }

\author{Xiao-Bin Liang}
\affiliation{School of Mathematics and Computer science, Shangrao Normal University,
 Shangrao 334001, China}
\author{Bo Li}
\email{libobeijing2008@163.com}
\affiliation{School of Mathematics and Computer science, Shangrao Normal University,
 Shangrao 334001, China}
\author{Shao-Ming Fei}
 \affiliation{Max-Planck-Institute
for Mathematics in the Sciences, 04103 Leipzig, Germany}\affiliation{School of Mathematical Sciences, Capital Normal University, Beijing 100048, China}

\begin{abstract}

In a recent paper, M. F. Sacchi [Phys. Rev. A 96, 042325 (2017)] addressed the general problem of approximating an unavailable quantum state by the convex mixing of different available states. For the case of qubit mixed states, we show that the analytical solutions in some cases are invalid. In this Comment, we present complete analytical solutions for the optimal convex approximation. Our solutions can be viewed as correcting and supplementing the results in the aforementioned paper.


\end{abstract}
\maketitle

In Sec. III of Ref. \cite{Sacchi15}, the problem of optimally generating a desired quantum state $\rho$  by the given set of the eigenstates of all
Pauli matrices was provided. Namely, consider the optimal convex approximation of a quantum state with respect to the set
\begin{align}\label{b3}
& B_3=\{|0\rangle,|1\rangle,|2\rangle=\frac{\sqrt{2}}{2}(|0 \rangle +|1 \rangle),|3\rangle=\frac{\sqrt{2}}{2}(|0 \rangle - |1 \rangle),\nonumber\\
& |4\rangle=\frac{\sqrt{2}}{2}(|0 \rangle + \sqrt{-1} |1 \rangle),|5\rangle=\frac{\sqrt{2}}{2}(|0
\rangle - \sqrt{-1} |1 \rangle)\}.\nonumber
\end{align}
The optimal convex approximation of $\rho$ with respect to $B_3$ is defined as $D_{B_3}(\rho)=min\{\parallel\rho-\sum_i p_i\rho_i\parallel_1\}$, where $\rho_i=|i\rangle\langle i|$,
$0\leq p_i\leq 1$, $\sum_i p_i=1$, the minimum is taken over all possible probability weights $\{p_i\}$, and $\parallel A\parallel_1$ denotes the trace norm of $A$, that is,
$\parallel A\parallel_1=Tr\sqrt{A^\dag A}=\sum_is_i(A)$ with $\{s_i(A)\}$ representing the singular values of $A$.
The optimal convex approximate set is given by $S(\rho^{opt})=\{\rho^{opt}|D_{B_3}(\rho)=\parallel\rho-\rho^{opt}\parallel_1\}$.

Here we point out that the analytical solution given in [Phy. Rev. A 96, 042325(2017)] is invalid in some cases.
We first provide a simple example. Consider the target qubit $\rho$ given by
\begin{equation}\label{taget}
\rho =\left(
\begin{array}{cc} 1-a
& k
\sqrt{a(1-a)}e^{-i\phi }\\
k \sqrt{a(1-a)}e^{i\phi }
& a  \\ \end{array} \right ) \;
\end{equation}
with $a \in [0,1]$, $\phi \in [0,2\pi]$, and $k\in [0, 1]$. If we set $a=1/2$, $k=1$, $\phi=\pi/4$, it is easily verified that the point belongs to the
region of case (i) in Ref. \cite{Sacchi15}, that is, $k_{th}\equiv a/({\sqrt{a(1-a)}(\cos\phi+\sin\phi)})< k\leq a/({\sqrt{a(1-a)} })$. Then the optimal convex approximation
and the corresponding optimal weights are given by Eq. (18) and (19) in \cite{Sacchi15}, respectively. However, if one substitutes $a=1/2$, $k=1$ and $\phi=\pi/4$ into Eq. (19) in \cite{Sacchi15}, one has $p_0=1-4a/3-2k\sqrt{a(1-a)}(\cos\phi+\sin\phi)/3=(1-\sqrt{2})/3<0$, which implies that the optimal probability is negative and this solution is invalid.

In the following, in terms of the method used in \cite{liang16} (see also the Karush-Kuhn-Tucker theorem and its conclusion in \cite{Forst}, p46-60),
we provide the complete analytical solution for the optimal convex approximation of a quantum state under $B_3$ distance and the corresponding optimal weights.

For simplicity, we denote $u=k\sqrt{a(1-a)}\cos \phi$, $v=k\sqrt{a(1-a)}\sin \phi$, where $k\in[0,1]$, $a\in[0,\frac{1}{2}]$ and $\phi\in[0,\pi/2]$.
When $a-u-v\geq 0$, one has $D_{B_3}(\rho)=0$. The pertaining weights corresponding to $\rho_i$ are given by
\begin{eqnarray}\label{protaget}
&&p_0=1-a-u-v -t_1-t_2,\nonumber
\\& &
p_1=a-u-v -t_1-t_2,   \nonumber \\& &
p_2=2u+t_1,\nonumber \\& &
p_3=t_1,\nonumber \\& &
p_4=2v+t_2,\nonumber \\& &
p_5=t_2,
\end{eqnarray}
where $t_1$ and $t_2$ are arbitrary non-negative arguments such that $p_1\geq0$.
If   $t_1=t_2=0$, then Eq. (\ref{protaget}) reduce to Eq. (14) in Ref. \cite{Sacchi15}.
However, if one sets  $t_1=a-u-v,t_2=0$ in (\ref{protaget}), one gets $p_0=1-2a$, $p_1=0$, $p_2=a+u-v$, $p_3=a-u-v$, $p_4=2v$ and $p_5=0$.
This is another kind of decomposition which is different from the one in Ref. [1]. Thus, our decompositions can be viewed as a complete supplement to the
results in Ref. \cite{Sacchi15}.

The previous complete analytical solution can be classified into the following four cases, see proof in Supplemental Material:

\par \noindent  $i)$
If $a<u+v\leq(3-4a)/2$, $a-v+2u\geq0$ and
$a-u+2v\geq0$, the optimal convex approximation of $\rho$ is given by
\begin{eqnarray}
D_{B_3}(\rho)= \frac{\sqrt{3}}{3}(\langle\sigma_x\rangle+\langle\sigma_y\rangle+\langle\sigma_z\rangle-1)
\;,\nonumber
\end{eqnarray}
with corresponding optimal weights
\begin{eqnarray}
&&p_0=1-4a/3-2u/3 -2v/3, \nonumber
\\& &
p_2=2a/3-2v/3+4u/3, \nonumber \\& &
p_4= 2a/3-2u/3+4v/3, \nonumber \\& &
p_1=p_3=p_5=0.  \nonumber
\end{eqnarray}

\par \noindent  $ii)$
If $a<u+v\leq(3-4a)/2$, $a-v+2u\geq0$ and $a-u+2v<0$, the optimal convex approximation of $\rho $ is given by
\begin{eqnarray}
D_{B_3}(\rho)= \sqrt{\langle\sigma_y\rangle^2+\frac{1}{2}(\langle\sigma_x\rangle+\langle\sigma_z\rangle-1)^2}
\;,\nonumber
\end{eqnarray}
with the corresponding optimal weights
\begin{eqnarray}
&&p_0=1-a-u, \nonumber
\\& &
p_2=a+u,\nonumber \\& &
p_1=p_3=p_4=p_5=0.   \nonumber
\end{eqnarray}

\par \noindent  $iii)$ If $a<u+v\leq(3-4a)/2$, $a-v+2u<0$ and $a-u+2v\geq0$, the optimal convex approximation of $\rho $ is given by
\begin{eqnarray}
D_{B_3}(\rho)= \sqrt{\langle\sigma_x\rangle^2+\frac{1}{2}(\langle\sigma_y\rangle+\langle\sigma_z\rangle-1)^2}
\;.\nonumber
\end{eqnarray}
The related optimal weights are given by
\begin{eqnarray}
& &p_0=1-a-v, \nonumber\\
& &p_4=a+v,\nonumber \\
& &p_1=p_2=p_3=p_5=0.\nonumber
\end{eqnarray}

\par \noindent  $iv)$ If $u+v>(3-4a)/2$, we have
\begin{eqnarray}
D_{B_3}(\rho)= \sqrt{\langle\sigma_z\rangle^2+\frac{1}{2}(\langle\sigma_y\rangle+\langle\sigma_x\rangle-1)^2}\nonumber
\end{eqnarray}
with the pertaining optimal weights
\begin{eqnarray}\label{e3}
&&p_2=1/2+u-v,\nonumber
\\& &
p_4=1/2-u+v,\nonumber \\& &
p_0=p_1=p_3=p_5=0.
\end{eqnarray}

\begin{figure}[!htbp]
\centering
\includegraphics[scale=0.4]{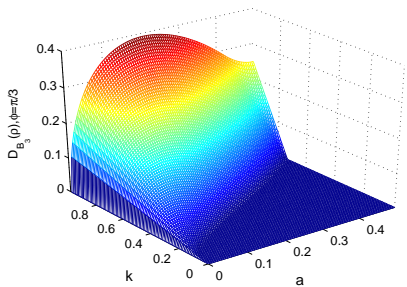}
\caption{ $D_{B_3}(\rho)$ as a function for the parameter $\phi=\frac{\pi}{3}$ from our result.}\label{fig_1}
\end{figure}

Up to now, we have refined the conclusions in Sec. III of Ref. \cite{Sacchi15}. Particularly, we have added the case $iv)$ as a valid supplement.
Moreover, we point out that the Fig. 2 in Ref. \cite{Sacchi15} is inaccurate in some areas. In the following, we plot the accurate $D_{B_3}(\rho)$ for fixed value of the phase parameter $\phi=\frac{\pi}{3}$, see Fig. \ref{fig_1}.

\begin{figure}[!htbp]
\centering
\includegraphics[scale=0.4]{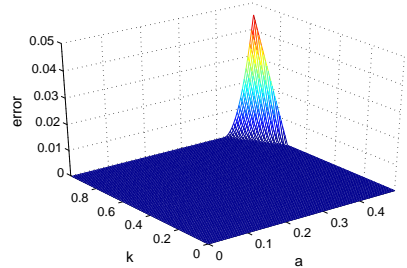}
\caption{The difference of $D_{B_3}(\rho)$ between our result and that of Ref. \cite{Sacchi15}, as a function of the phase parameter $\phi=\frac{\pi}{3}$.}\label{fig_2}
\end{figure}

As another related example, consider $k=1$ and $a=1/2$. According to Eq. (18) in Ref. \cite{Sacchi15},
we get that $D_{B_3}(\rho)$ is about 0.2113. From Eq. (19) of \cite{Sacchi15}, $p_0 =(\sqrt{3}-1)/3 <0$.
In fact, according to (\ref{e3}), the accurate $D_{B_3}(\rho) \approx0.2588$, and the corresponding probability is $p_0=p_1=p_3=p_5=0$, $p_2=(1+\sqrt{3})/4$ and $p_4=(3-\sqrt{3})/4$.
We plot the difference of Fig. \ref{fig_1} in our paper and Fig. 2 in \cite{Sacchi15}, see Fig. \ref{fig_2}.


In summary, we have derived the complete solution for the optimal convex approximation of a qubit mixed state under $B_3$ distance. We have revised the problem related to the result for $a<u+v$ in Ref. \cite{Sacchi15}. In addition, if $a\geq u+v$, our decompositions are the complete supplement to the representative decompositions in Ref. \cite{Sacchi15}.

We would like to say that the idea of looking for the least distinguishable states is nice,
and the condition Eq. (13) in Ref. \cite{Sacchi15} for exact convex decomposition is also correct.
We would also point out that the discussion in the last section of Ref. \cite{Sacchi15}, on the case of many copies of quantum states, the non-additivity of the distance, and the role of correlations, maintain general validity.


\section{Appendix}\label{Appendix}
We now provide a detail proof of the state classification in the main text.
To find $D_{B_3}(\rho)=min\parallel\rho-\sum_i p_i\rho_i\parallel_1$ is
equivalent to search for the solution of the following minimum,
\begin{eqnarray}
Min \{2\sqrt{\mid Det(\rho-\sum _{i}  p_i|i\rangle\langle i| )\mid}\},
\end{eqnarray}
such that $p_i\geq 0$ and $\sum _{j}p_j=1$. Denote
\begin{align}\label{question}
f(p_0,p_1,p_2,p_3,p_4,p_5)&=&\mid Det(\rho-\sum^5 _{i=0}  p_i|i\rangle\langle i|  ) \mid \nonumber\\
&&-\sum^5 _{i=0}\lambda_i p_i-\lambda\sum^5 _{i=0}p_i.
\end{align}
According to the Karush-Kuhn-Tucker Theorem \cite{Forst} and the related conclusion (see page 46-60 in \cite{Forst}),
the above question is equal to
\begin{align}\label{question1}
\nabla f=0,~\lambda_ip_i=0,~\lambda_i\geq0,~p_i\geq0,~\sum^5 _{j=0}p_j=1
\end{align}
for $i=0,1,2,3,4,5$.
One then obtains the following equations and inequalities
\begin{eqnarray}\label{question2}
&&p_1+p_2/2+p_3/2+p_4/2+p_5/2+\lambda_0+\lambda-a=0, \nonumber
\\& &
p_0+p_2/2+p_3/2+p_4/2+p_5/2+\lambda_1+\lambda-1+a=0,\nonumber \\& &
p_0/2+p_1/2+p_3+p_4/2+p_5/2+\lambda_2+\lambda-1/2+u,\nonumber \\& &
p_0/2+p_1/2+p_2+p_4/2+p_5/2+\lambda_3+\lambda-1/2-u=0,\nonumber \\& &
p_0/2+p_1/2+p_2/2+p_3/2+p_5+\lambda_4+\lambda-1/2+v=0,\nonumber \\& &
p_0/2+p_1/2+p_2/2+p_3/2+p_4+\lambda_5+\lambda-1/2-v=0,\nonumber \\& &
\lambda_ip_i=0,~\lambda_i\geq0,~p_i\geq0,~i=0,1,2,3,4,5,\nonumber \\& &
\Sigma _ip_i=1,
\end{eqnarray}
where $u=k\sqrt{a(1-a)}\cos \phi$ and $v=k\sqrt{a(1-a)}\sin \phi$. From (\ref{question2}) we have

(1) If $p_0\neq0, p_1\neq0$, from $\lambda_1=\lambda_2=0$ we have
$\lambda=0$ and $\lambda_i=0$ ($i=2,3,4,5$). Similarly, if $p_2\neq0,~ p_3\neq0$ or $p_4\neq0,~ p_5\neq0$ or at least four of $\{p_i\}$ are nonzero, (\ref{question2})
is equivalent to $\nabla f=0,$ $\lambda=0,$ $\lambda_i=0$ ($i=0,1,2,3,4,5$), $\Sigma _ip_i=1$. Thus we have
\begin{eqnarray}
&&p_0=1-a-u-v -t_1-t_2,\nonumber
\\& &
p_1=a-u-v -t_1-t_2,   \nonumber \\& &
p_2=2u+t_1,\nonumber \\& &
p_3=t_1,\nonumber \\& &
p_4=2v+t_2,\nonumber \\& &
p_5=t_2,\nonumber
\end{eqnarray}
where $t_1$ and $t_2$ are arbitrary non-negative numbers such that $p_1\geq0$. In this case, $D_{B_3}(\rho)=0$,
and the condition $p_i\geq0$, $i=0,1,2,3,4,5$, is transformed to $a-u-v\geq 0$.

(2) Only three numbers of $\{p_i\}$ are nonzero. According to (1), $D_{B_3}(\rho)=0$ for $a-u-v\geq 0$. For $a-u-v< 0$,
that only three numbers of $\{p_i\}$ are nonzero results in the following cases,
$(i)$: $p_0\neq0$, $p_2\neq0$, $p_4\neq0$; $(ii)$: $p_0\neq0$, $p_2\neq0$, $p_5\neq0$;
$(iii)$: $p_0\neq0$, $p_3\neq0$, $p_5\neq0$; $(iv)$: $p_1\neq0$, $p_2\neq0$, $p_4\neq0$;
$(v)$: $p_1\neq0$, $p_2\neq0$, $p_5\neq0$;
$(vi)$: $p_1\neq0$, $p_3\neq0$, $p_5\neq0$.
However, the solution of (\ref{question2}) exists only for the case $(i)$, that is,
\begin{eqnarray}
&&p_0=1-4a/3-2u/3 -2v/3, \nonumber
\\& &
p_2=2a/3-2v/3+4u/3, \nonumber \\& &
p_4= 2a/3-2u/3+4v/3, \nonumber \\& &
p_1=p_3=p_5=0.  \nonumber
\end{eqnarray}
We obtain that $a<u+v\leq(3-4a)/2$, $a-v+2u\geq0$ and $a-u+2v\geq0$.

Now we illustrate that for the case $(iv)$, no solution exists. According to the assumption, $\lambda_1=0$, $\lambda_2=0$ and $\lambda_4=0$.
Then $p_1=4a/3-1/3-2u/3-2v/3$, $p_2= 2/3-2a/3+4u/3-2v/3$ and $p_4=2/3-2a/3-2u/3+4v/3$.
Notice that $p_1=4a/3-1/3-2u/3-2v/3\geq0$. One obtains $a<u+v\leq2a-1/2$, namely,
$a>1/2$, which results in a contradiction. One can similarly show that for the cases $(ii)$, $(iii)$, $(v)$ and $(vi)$, also no solutions exist.

(3) Now consider the region $a<u+v\leq(3-4a)/2$.

Case $i$: $a<u+v\leq(3-4a)/2$, $a-v+2u\geq0$ and $a-u+2v<0$.
For  $p_0\neq0$ and $p_2\neq0$, the equation (7) has the following solution,
\begin{eqnarray}
&&p_0=1-a-u, \nonumber
\\& &
p_2=a+u,\nonumber \\& &
p_1=p_3=p_4=p_5=0.   \nonumber
\end{eqnarray}

Case $ii$: $a<u+v\leq(3-4a)/2$, $a-v+2u<0$ and $a-u+2v\geq0$, that is, $p_0\neq0$ and $p_4\neq0$, the  solution of (7) is given by
\begin{eqnarray}
& &p_0=1-a-v, \nonumber
\\& &
p_4=a+v,\nonumber \\& &
p_1=p_2=p_3=p_5=0.\nonumber
\end{eqnarray}

(4) The last case, $u+v>(3-4a)/2$ and only two numbers of $\{p_i\}$ are nonzero.
In this case, only for $p_2\neq0$ and $p_4\neq0$, one has the following solution,
\begin{eqnarray}
&&p_2=1/2+u-v,\nonumber
\\& &
p_4=1/2-u+v,\nonumber \\& &
p_0=p_1=p_3=p_5=0.\nonumber
\end{eqnarray}
For the other 6 cases with only two nonzero $\{p_i\}$, there do not exist solutions.
For example, let us consider the case $p_1\neq0$ and $p_2\neq0$.
By $\lambda_1=0$ we have $\lambda\leq 0$. Thus, $p_1=a-u$, $p_2=1-a+u$, $\lambda=1/2-a/2-u/2$, which leads to a contradiction, $1-a\leq u\leq a$ and $p_1=0$.

\bigskip
\noindent {\bf Acknowledgments}
This work is supported by NSFC (11765016,11675113), Jiangxi Education Department Fund (KJLD14088) and Key Project of Beijing Municipal Commission of Education under No. KZ201810028042.

\end{document}